# Understanding and Designing Automation with Peoples' Wellbeing in Mind

**Holger Klapperich**
`Ubiquitous Design´, Experience & Interaction
University Siegen, Germany
holger.klapperich@uni-siegen.de

**Marc Hassenzahl**
`Ubiquitous Design´, Experience & Interaction
University Siegen, Germany
marc.hassenzahl@uni-siegen.de

**Alarith Uhde**
`Ubiquitous Design´, Experience & Interaction
University Siegen, Germany
alarith.uhde@uni-siegen.de

## ABSTRACT

Nowadays, automation not only dominates industry but becomes more and more a part of our private, everyday lives. Following the notion of increased convenience and more time for the "important things in life", automation relieves us from many daily household chores – robots vacuum floors and automated coffeemakers produce supposedly barista-quality coffee on the press of a button. In many cases these offers are embraced by people without further questioning. Of course, automation frees us from many unloved activities, but we may also lose something by delegating more and more everyday activities to automation. In a series of four studies, we explored the experiential costs of everyday automation and strategies of how to design technology to reconcile experience with the advantages of ever more powerful automation.

**KEYWORDS**

Home automation; coffee brewing; design for wellbeing, meaningful experiences, automation from below

**INTRODUCTION**

Since the Stone Age, humans have been creating tools such as hand axes from flint stone to cut meat. Over the centuries, these tools became more and more powerful and efficient, allowing for new abilities as well as easing the strains of physical work in many different ways. In the second half of the 18th century, toolmaking gradually turned into automation, with spinning mules and looms powered by horses, water, or later steam. These tools produced goods, such as thread, with minimal physical labour. While labour got transferred from humans to machines, this was not primarily done to spare humans hard work, but to produce goods more efficiently and profitably. Automation became the foundation of modern capitalism and got deeply ingrained into visions of modernity. It thus comes as no surprise that, beginning with the early 20th century, automation found its way from shop floors into homes. For example, the fully automated toaster supports the busy housewife since 1926. With automation, related values entered the home. Efficiency had entered everyday private life as much as business life.

Everyday automation was and is predominantly driven by technical feasibility rather than by the visions of creating a fulfilling cooperation between humans and automation. While from the perspective of today, the toaster is a welcome shortcut compared to roasting bread on an oven, we are certain that some may have missed the activity – for example, because they had been especially skillful bread roasters. They may have enjoyed the resulting morning ritual or loathe that the time saved is now occupied by the less inspiring tasks of cleaning the toaster.

While automation has many advantages, it is similarly well known that automation has a number of negative side effects, such as alienation, deskilling, and overreliance. Aporta and Higgs [1], for example, studied how the Inuit lost their skills in wayfinding after GPS-technology became widely used. Originally, wayfinding skills were passed from generation to generation. This stopped with the rise of navigation systems. As a consequence, the Inuit and their survival under harsh conditions became highly dependent on GPS, since they could not fall back to practiced skills and intuition. In another example, Sheridan and Parasuraman [8] described deskilling and the effect of being out-of-the-loop when piloting highly automated aircrafts. They argue not to forget that the "[…] ultimate purpose of technology is to make life better for people."

We argue that everyday automation brought problems, such as alienation, deskilling and overreliance, from the shop floor into our homes. While the modernist narrative of the ever more powerful automation understands "better life" in terms of getting rid of supposedly unwanted chores, a more humanist perspective may question this oversimplification. To us, a "better life" does not imply better technology per se but to foster people's wellbeing by designing positive, that is, enjoyable and meaningful, everyday experiences [4].

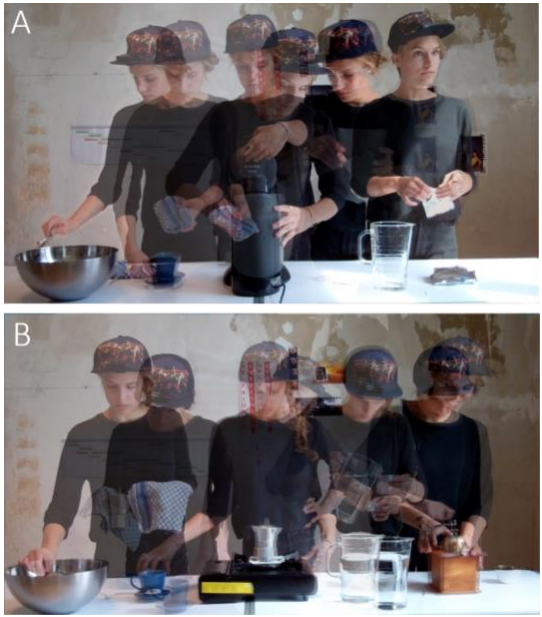

**Figure 1: (A) Automated and (B) manual setup**

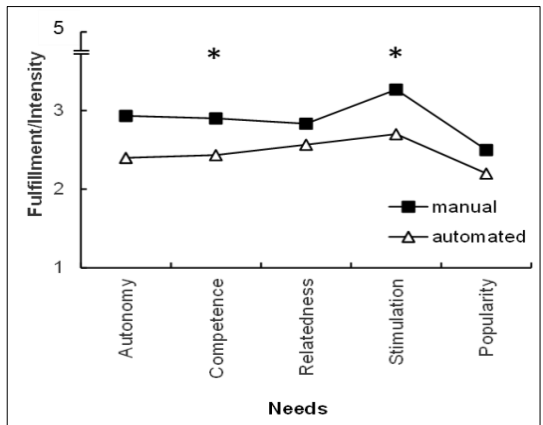

**Figure 2: Mean fulfillment for manual and automated coffee brewing**

More specifically, an enjoyable and/or meaningful experience can be understood as a moment, where psychological needs such as competence, autonomy, security, popularity, relatedness, competence, stimulation, and physical thriving, become satisfied through an activity. Since the material (e.g., things, tools, devices) shapes activities and experiences, automation will inevitably impact the quality of everyday experiences. In addition, since automation is "designable", diverse forms of automation may impact experiences in different ways.

In the present paper, we outline four studies into the question of whether everyday automation in its current form is increasing meaning or rather decreasing it and how automation could be designed to become meaningful.

**THE EXPERIENTIAL COSTS OF EVERYDAY AUTOMATION**

In a first study [5], we took a look at the differences in emerging experiences between automated coffee brewing (with a Senseo pad machine) and more manual coffee brewing (with a manual grinder in combination with an Italian coffee-pot / Fig.1). Participants (N=20) had to prepare two cups of coffee in either way. After each cup we asked for the most positive and most negative moment during the activity. In addition, participants were asked to fill in a questionnaire, measuring positive and negative affect as well as psychological need satisfaction, such as autonomy, competence, popularity, stimulation, security, and relatedness.

Overall, brewing coffee manually was rated as more positive and more need fulfilling than brewing it in an automated way. Feelings of competence and stimulation were significantly higher when performing the "manual process" (Fig.2). However, the manual process also led to more negative affect and took thrice as long. The dilemma of automation become obvious. While automation is clean, efficient, and convenient, the meaningful moments are lost, since automation makes activities experientially "flat" in terms of need fulfillment and positive affect. Performers of the manual brewing enjoyed the smell, haptics, and the sensory stimulation while grinding. They felt as a part of the process through the transparency of each single step. They also felt in control and competent. While this engagement could lead to positive as well as negative moments, automation simply made the whole activity more or less disappear. Quite tellingly, 19 out of 20 people mentioned the waiting time as the most negative moment when brewing automatically, an aspect which did not play any role when manually brewing although this took thrice as long. In sum, this study showed that different ways of performing an activity (i.e., making coffee) leads to different experiential consequences. While automation is efficient and makes impatient, a more manual way provides more positive affect and more need fulfillment, i.e., meaning.

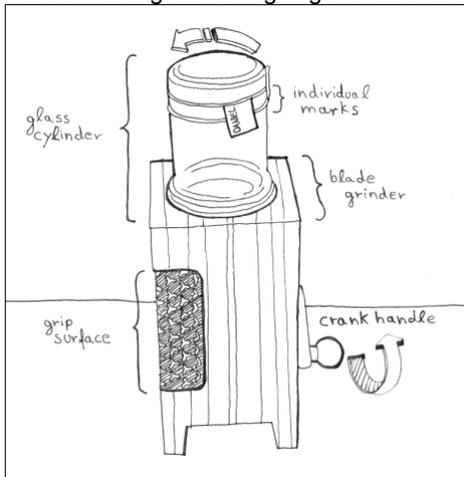

**Figure 3: Elements of Hotzenplotz**

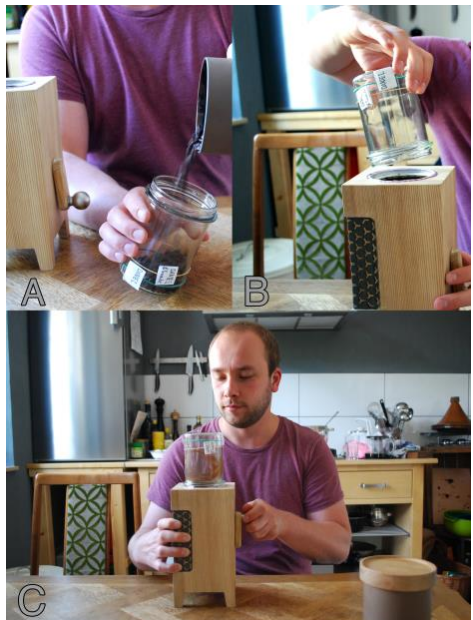

**Figure 4: Hotzenplotz in action**

**IMPROVING WELLBEING THROUGH MANUAL COFFEE PREPARATION IN EVERYDAY LIFE**

The first study implies that a more manual way of doing an activity could be a source for wellbeing and meaning. However, the study took place in the lab, where participants had dedicated time for the activity. It seems probable that introducing manual brewing in everyday life will create more negative feelings due the time and effort needed. We explored this by introducing households, who typically prepare coffee automatically, to an additional, more manual way. All households (N=10 [number of households], n=25 [number of people]) owned a coffee machine that was used in everyday life. In the first week, the participants continued to use their regular coffee maker. We asked them to fill in a questionnaire (an abbreviated form of the ones used in the first study) each time they made coffee (baseline). In the second week, each household was provided with a manual preparation kit, consisting of a French-Press coffee maker, a manual coffee bean grinder, and a bag of coffee beans. The beans could be used for both the manual and automated coffee preparation. It was left to the household members how to prepare their coffee, although we encouraged all people in the households to try the manual preparation at least once. Participants where again asked to fill in a short questionnaire after each preparation, both manual and automated.

Overall, in the baseline condition all participants engaged in 273 instances of coffee preparation. In the second week, they used the manual way 106 times and the automated way 93 times. Comparing the baseline week and the manual condition in week 2, the manual process led to more positive affect. We further compared the baseline with automated brewing in week 2 and found that, after the manual alternative was introduced, automatic brewing was experienced as less stimulating. When comparing the manual and automated condition in week 2 after the French press was introduced, we found that manual coffee making created more feelings of stimulation and competence and more positive affect. All in all, a manual way of brewing coffee introduced to households on a voluntary basis was taken up in a little more than 50% of all instances of coffee brewing. In line with the first study, manual brewing led to more positive affect and need fulfillment.

**RECONCILING AUTOMATION WITH EXPERIENCE THROUGH INTERACTION DESIGN**

Obviously, what we labeled as "manual" brewing in studies 1 and 2 was not strictly manual. It also made use of tools, such as a grinder, however, affording different interactions and yet shaping resulting activities and experiences differently. In the third study we set out to deliberately design an interaction for an electric grinder, which preserves experiential quality and thus reconciles automations with experience. *Hotzenplotz* [6] (Fig.3) was designed based on a concept we dubbed "automation from below". This kind of interaction starts from a manual interaction, which however becomes supported by automation. An everyday example would be an electric bike, which just supports pedaling but does not replace it. We fitted an electric grinder with a crank.

A rotation sensor inside the crank detected turning, according signals are forwarded to an Arduino board. A simple program then controls the speed of the motor of the grinder depending on the turning motion: Slow turning led to a slow rotation of the grinder's two blades; fast turning accelerated the blades. In other words, *Hotzenplotz* borrowed an archaic, yet meaningful interaction, to presumably inject meaning into the activity of switching on an electrical grinder.

In a lab study, we compared *Hotzenplotz* with a manual grinder and an electric grinder (of the same model used in *Hotzenplotz*). Participants had to grind 3 tablespoons of beans with each grinder. Subsequently, they filled in a number of questionnaires. As expected, manual and combined *Hotzenplotz*-style grinding were experienced as more positive and more meaningful than electric (automated) grinding. In fact, on some measures, such as positive affect and hedonic quality, combined grinding was experienced even slightly better than manual grinding. Interestingly, combined grinding was also perceived as more pragmatic than the electric and the manual grinder. Concerning time perception, participants wanted to shorten the electric grinding, while manual grinding was balanced. With *Hotzenplotz*, however, participants wanted to prolong the interaction.

All in all, while *Hotzenplotz* had all the advantages of an electric grinder (e.g., finer powder) involving a presumably "manual" interaction, created positive experiences and meaning through interaction. Contrary to a notion of "form follows function", a technically superfluous crank was able to instill meaning typically lost when using an electric grinder. This is an admittedly simple but nevertheless interesting example of reconciling automation with experience.

**Wellbeing-oriented design of automation in the field**

The previous study was lab based. It is possible, that *Hotzenplotz* is stimulating and interesting in a brief lab interaction, but will not create much change in perception and experience when used in everyday life. To explore this, we provided eight participants who typically brew their coffee "manually", with *Hotzenplotz* and asked them to use the respective grinder in their daily routines. Through interviews, we explored the effects of *Hotzenplotz* on wellbeing. In the following, we provide some examples.

When using the electric grinder, P2 felt out of the loop: "I thought I'd like to have fresh coffee powder and the machine provides it. But I'm not needed for it […] that's the reason why it is meaningless to me." P4 descried this even more drastically: "the process definitely got lost." At the same time P4 also saw advantages: "I always did something in parallel. I have not observed it [the grinding process]." Most participants reported the advantage of increased efficiency with the electric grinder, but the disadvantages of feeling out of control and missing the meaning of grinding their coffee beans. When using the *Hotzenplotz*, participants felt back in the loop. P1 stated "I liked participating by turning the crank. I could control the volume [of the grinder]. And I liked to have a look at it."

P2 described the effect of "automation from below" in more detail "It was fun when I started turning the crank […] and suddenly it [*Hotzenplotz*] took over, but I still was in charge to keep on turning the crank." Moreover, P2 praised the sensitive automation "You won't expect such an interaction, because if you use a manual coffee grinder it is super exhausting […], here it was surprisingly easy and that was fun." P2 further stated that this was also because of the transparency of the system: "through the glass cylinder you could see the coffee beans flying around, you recognize the grinding degree […] and you can stop or continue it, that is positive." Also, P3 described the same phenomena but added: "I made time [for the grinding] consciously." She also described the process: "I had the feeling that the machine and I cooperated." All in all, using *Hotzenplotz* in the field created quite positive effects, comparable to our lab study.

**Designing for synergetic effects of manual and automated processes**

Automation has advantages, such as enhancing the capabilities of the user or saving time. However, it also bears the risk of removing a potential for happiness in everyday life. In a worst-case scenario, automation could lead to boredom, deskilling, and alienation. However, this should not be a reason to avoid automation, but to design it experientially rich and thus meaningful, while keeping the advantages. With *Hotzenplotz* we explored "automation from below". However, other strategies are possible, such as making the automation more experienceable per se, that is, to create feelings of involvement in a "user" without being actually involved. A further example is Grosse-Hering et al.'s [2] strategy to deliberately prolong and emphasize meaningful moments of an activity and to downplays less important aspect. Interestingly, modern automated coffee machines embody the opposite approach. The meaningful elements, such as brewing the coffee, grinding the beans, is done by the machine, while the user is needed to do mainly cleaning and maintenance (refilling water, emptying the drip tray).

For the present studies, we used coffee making as a sort of "model" activity, but we are convinced that our findings transfer to other domains, such as automated driving [2].

We need to be more aware of the consequences of everyday automation and to develop design strategies to reconcile automation with experience. From a technological perspective, fitting an electric grinder with a crank may appear absurd. From a wellbeing-oriented, experiential perspective certainly not. Wellbeing-oriented design methodology [7] might help to guide the design of automation in a more sensible way. Future research is needed to find a way how designers can deliberately create subjective wellbeing though automation.